\documentclass[reprint,aps,jcp,amsmath,amssymb,superscriptaddress,floatfix,showpacs,twocolumn]{revtex4-1}

\usepackage{amsmath}
\usepackage{amssymb}
\usepackage{graphicx}
\usepackage{dcolumn}
\usepackage{bm}

\usepackage{ftnxtra}

\input epsf.sty

\usepackage{color}
\newcommand{\todo}[1]{}
\renewcommand{\todo}[1]{{\color{red} TODO: {#1}}}

\newcommand{\done}[1]{}
\renewcommand{\done}[1]{{\color{blue}  Done: {#1}}}
\makeatletter
\def\captionof#1#2{{\def\@captype{#1}#2}}
\makeatother

\begin{document}
\title{The Early Crystal Nucleation Process in Hard Spheres shows Synchronised Ordering and Densification}

\author{Joshua T.~Berryman}
\affiliation{Theory of Soft Condensed Matter, Universit\'{e} du Luxembourg, L-1511 Luxembourg, Luxembourg}
\author{Muhammad Anwar}
\affiliation{Theory of Soft Condensed Matter, Universit\'{e} du Luxembourg, L-1511 Luxembourg, Luxembourg}
\affiliation{Department of Mechanical Engineering, Institute of Space Technology, Islamabad, Pakistan}
\author{Sven Dorosz}
\author{Tanja Schilling}
\affiliation{Theory of Soft Condensed Matter, Universit\'{e} du Luxembourg, L-1511 Luxembourg, Luxembourg}

\begin{abstract}
We investigate the early part of the crystal nucleation process in the hard sphere fluid using data produced by computer simulation. We find that hexagonal order manifests continuously in the overcompressed liquid, beginning approximately one diffusion time before the appearance of the first `solid-like' particle of the nucleating cluster, and that a collective influx of particles towards the nucleation site occurs simultaneously to the ordering process: the density increases leading to nucleation are generated by the same individual particle displacements as the increases in order. 
We rule out the presence of qualitative differences in the early nucleation process between medium and low overcompressions, and also provide evidence against any separation of translational and orientational order on the relevant lengthscales.
\end{abstract}

\maketitle

\section{Introduction}
As a typical first-order phase transition, crystallization from the metastable melt begins with 
a nucleation process. There are two order parameters which 
characterize the transition (density and `crystalline order'). 
It is not {\it a priori} evident that both parameters 
undergo the same dynamics during the transition process. Consequently, 
a ``density-first'', a ``bond-order-first'', and
several more complex phase transition scenarios have been proposed and 
vividly discussed in the literature.
\\

In this article we investigate the role of density and order fluctuations prior to nucleation and in the environment of the growing nucleus. 
As a model system we use spherical particles that have repulsive interactions only. 
A hard sphere model is the starting point for many theoretical treatments of granular, fluid, glassy and crystalline systems, and may be sufficient without further refinement if excluded volume interactions are more significant than long-range forces. Technological examples where a hard sphere model is sufficient within important regimes include the study of metal solidification (e.g.~\cite{Zhang2015}) and the formation of colloidal crystals (e.g.~\cite{Cheng1999}).
Crystallization in hard spheres has been studied extensively but retains many open questions, of which the nature of the initiating fluctuation is a particularly active concern.
\\

Classical Nucleation Theory (CNT) assumes that the emerging nucleus and the surrounding fluid possess the properties of the respective bulk phases, i.e.~that the emerging nucleus already has the order and density of the bulk crystal and that the surrounding fluid is not affected by the growth of the nucleus. This picture is not immediately credible in real liquids and colloidal suspensions, where  transport of material is relevant. A detailed consideration of the outcome of mass conservation and finite transport speeds in the fluid has been made for vapour-liquid nucleation by Lutsko et al.~\cite{Lutsko2006,Lutsko2011,Lutsko2012}, leading to an expectation of substantial early density changes on a lengthscale larger than the initial nucleus (for the vapor-liquid nucleation of Lennard-Jones particles). Early densification in fluid-solid transitions is not {\it a priori} expected to be as important as in vapour-liquid transitions, due to the smaller difference in densities of the two phases, however this phenomenon has been reported in computer simulations of hard sphere crystallisation by Schilling et al.~in 2010 \cite{Schilling2010} as well as from experiment \cite{Schoepe2006, Zhang2007}.
\\

A computational study by Russo \& Tanaka \cite{Russo2012} has examined structure and density changes for a set of nucleating trajectories at a number density $\phi\sigma^3\approx 1.02$, where $\sigma$ is the diameter of a particle. Augmenting these trajectories with fluctuation data drawn from the metastable liquid, they note a coupling of order and density, but state that ``the density increase is foreshadowed by the prestructuring of the nucleus'', adopting a position which we will crudely summarise as `order-first'.
\\

Tan et al.~presented, from optical microscopy, two distinct nucleation pathways for their colloidal system. Both pathways began with hexagonally-ordered precursors then developing into either bcc or fcc metastable structures.  
A three step process was described, with the final phase being either bulk bcc or rhcp depending on the radius of the particles.  The presence of bulk bcc, and also the dependence on radius, indicate that forces other than excluded volume were important in these experiments, however they remain an interesting reference in that the authors state that the nucleation sites were not correlated with increased local density. 
In fact the authors claim that ``nucleation rarely starts from the densest regions'' \cite{Peng2014}, see also a comment on this work by Gr\'an\'asy \& T\'oth \cite{Granasy2014}. 
Another investigation using microscopy, by Lu et al.~\cite{Yunzhou2015}, also found a decoupling of the density and the nucleation event, stating: ``nucleation events were observed that rarely start from the denser regions of colloidal samples''.
\\

Kawasaki \& Tanaka \cite{Kawasaki2010a} found medium range ordered precursors of sizes larger than the critical nucleus in simulations of hard spheres. The lifetime of these structures was estimated as ``a few times the relaxation time $\tau_\alpha$''. 
It was stated that the ordering was not icosahedral but rather hcp with multiple defects. A similar observation was reported by Schilling et al.~\cite{Schilling2010,Schilling2011} where low symmetry clusters of considerable size spontaneously transformed into highly ordered crystals.  
The effect of ordered precursor structures on crystallization should be to reduce the surface tension, which would be consistent with experiments of Gasser et al.~\cite{Gasser2001} finding strongly aspherical shapes for the critical or near critical nuclei.
The most recent minireview \cite{Granasy2014} states that fluid-solid nucleation is probably multistep in most cases, but that the specific steps expected to be relevant in a given system are to date unclear.
\\

The overview of the current literature is confusing: is the now-common idea of hard sphere nucleation as a two step process physically useful? 
Are densification and ordering separable from each other, and if so does one in general come first?  
The emergence of unclear and apparently contradictory statements creates a need to re-examine the onset of crystallization.
It is not desirable to reduce nuanced expositions from the literature to blunt 
and simplified statements, however we must remark that the 
`density first' \cite{Schilling2010} and `order first' \cite{Russo2012} positions cannot both be right.
The summarised claim in Schilling et al.~2010 \cite{Schilling2010} is that of a two step mechanism, remarkable in that it arises without attractive forces: ``The metastable fluid relaxes the density first, by producing dense low symmetry clusters''.  
The key claim from Russo \& Tanaka 2012 \cite{Russo2012}) is of the opposite two step mechanism ``The transition from liquid-like to crystal-like happens at constant density''. Kawasaki \& Tanaka argue in a separate work \cite{Kawasaki2010a} that the sequence in detail is {\it liquid} $\rightarrow$ {\it hcp} $\rightarrow$ {\it dense rhcp}.  
\\

To clarify the sequence of events we argue that, at least at low and medium overcompressions,  the fluid does indeed possess correlations over medium lengthscales in hexagonal order however these are typically accompanied by synchronised density fluctuations of the same lengthscale: the initiating fluctuation is of medium range and of low amplitude in both density and order, and has the same pattern of radial decay in each parameter as for the normal quiescent fluctuations.
Our analysis shows a synchronised increase of density and hexagonal order leading up to nucleation, although we allow that elevated hexagonal order is probably the more useful (less noisy) of the two reaction coordinates as a predictor of nucleation events.
\\ 

A further issue which has been mentioned in relation to this subject is the existence of two distinct ordered phases in 2D systems of hard discs.
The hard disc phase intermediate between the fluid and solid (called the hexatic) is distinct from the crystal in that is has long-range orientational order and short-range translational order \cite{Strandburg1988}. 
There is no known hexatic-like phase of bulk 3-spheres, however the possibility of some related unstable state manifesting along the nucleation pathway has generated a certain amount of excitement, therefore translational and orientational order are typically distinguished in the literature in case they should turn out to be meaningfully different to each other. 
\\

We argue that to claim a separation of translational and orientational ordering in 3 dimensional hard spheres would require much stronger evidence than has yet been seen: well-defined and distinct scaling of orientational and translational order in the fluctuations prior to nucleation events would need to be shown, and this has not been observed.  
Orientational order is conventionally defined with respect to the neighbourhood of a single particle via a decomposition of the orientational distribution of `bonds' (displacement vectors between neighbouring particles) into spherical harmonics ($q_6,y_6$ etc.), while translational order is typically computed using reciprocal-space observables not directly comparable to the orientational analysis. 
We attempt to test this distinction here by comparing translational and orientational order as directly as possible, via a direct-space analysis of bond lengths which is more comparable to the spherical harmonic analysis of bond orientations conventionally used to report the orientational order.
\\

 In this work we analyze new simulations at low volume fractions and using different dynamical schemes, and we also revisit the 2010 dataset. We show that local and also medium-range density changes take place simultaneously to the initial formation of (weak) translational and orientational order, presenting an order-with-density model which is distinct from the order-first mechanism preferred by Russo \& Tanaka \cite{Russo2012}. We also test the expectation based on work by Kawasaki et al.~\cite{Kawasaki2010a}, Tan et al.~\cite{Peng2014}, Barros \& Klein \cite{Barros2013}, and Schilling et al.~\cite{Schilling2010,Schilling2011} that weakly ordered precursors to crystallization should or might be present, and find that hexagonal ordering manifests gradually and continuously in the fluid prior to the formation of the first definitively solid-like particles, without evidence of any intermediate state.
\\

\section{Re-analysis and Extension of the 2010 Nucleation Dataset}

As outlined in the introduction, previous studies of simulation data have suggested an important role for dense precursors in the structure formation process of hard spheres \cite{Schilling2010}. We revisit an extant dataset comprising independent runs of $N=216000$ hard spheres of diameter $\sigma$ at a number densities of $\phi\sigma^3=1.03$, $1.027$ and $1.024$, corresponding to chemical potential differences in of about $|\Delta\mu|\simeq0.6k_BT$ per particle. Nucleation at these overcompressions was observable without the use of accelerated sampling algorithms. The time evolution was defined by MC simulation mimicking Brownian diffusion, with particles undergoing independent displacements at a fixed attempt frequency and small random stepsize much less than $\sigma$. In order to extend the original datasets, further runs were made, using event-driven molecular dynamics (MD) at $\phi\sigma^3=1.03$ thus testing for effects arising from the choice of dynamical scheme, and also using MD simulations accelerated by flat histogram Pruned and Enriched Rosenbluth Forward Flux Sampling (flatPERM-FFS) at lower overcompressions (tab.~\ref{tab:runlist}).\\

The known sampling difficulties for FFS calculations in this system \cite{Filion2010} were ameliorated using the flat histogram Pruned Enriched Rosenbluth FFS method (flatPERM-FFS) \cite{Rosenbluth1955,Prellberg2004,Allen2009} which was added to the functionality of the \verb|freshs.org| \cite{Kratzer2014} sampling system  for the purpose of this calculation.  flatPERM-FFS was found to give significant improvement in convergence relative to direct FFS. flatPERM-FFS is a path sampling technique applicable to stochastic dynamical systems subject to rare events, i.e.~bottlenecks in their dynamics such as crystal nucleation.  The scheme operates by selectively branching multiple copies of trajectories which make progress with respect to a specific collective variable (here the number of particles in the largest crystalline cluster).  As the dynamics are stochastic, the branched trajectories diverge with a proportion of them making further forward progress in the reaction coordinate.  The gist of the gain from using flatPERM rather than direct FFS is that paths are selected for branching in a Bayesian way that takes account of their histories.\\

\begin{table}[]
  \begin{center}
    \begin{tabular}{|l|l|r|r|c|c |}
     \hline
      $\phi\sigma^3$ & Scheme & N runs & N particles & $\Delta \mu_{|V}$ [$k_BT$] & Year\\
     \hline
      1.030 & MC  &  4 & 216000 & -0.585 & 2010\\
      1.030 & MD  &  6 & 216000 & -0.585 & new \\
      1.029 & flatPERM-FFS & 21 &  20262 & -0.58  & new \\
      1.027 & MC  &  6 &  64000 & -0.56  & 2010\\
      1.012 & flatPERM-FFS & 15 &  19924 & -0.50  & new \\
     \hline      
    \end{tabular}
    \caption{Simulations discussed in this work.  Runs using rare event sampling (flatPERM-FFS) were computationally cheaper than brute force MD and MC, therefore more runs were made (and smaller errorbars achieved).}
    \label{tab:runlist}
  \end{center}
\end{table}

Voronoi volumes were determined using Rycroft's \verb|voro++| tool \cite{Rycroft2009}. The crystalline cluster was identified by the dot product of Steinhardt bond order parameter $\vec{q}_6$ of neighbouring particles \cite{Steinhardt1983,tenWolde1996,Lechner2008}:

\begin{eqnarray}
q_6^m(i)       &=& \frac{1}{12} \sum_{j=1..12} Y_6^m(\vec{r}_{ij}) \\
q_6q_6(i,j)    &=& \frac{4\pi}{13}\sum_{m=-6..6} q_6^m(i)q_6^{m*}(j) 
\end{eqnarray}
 
where $Y_6^m(\vec{r})$ is the $m^{{\rm th}}$ of 13 complex components of the sixfold spherical harmonic, and where a bond was treated as solid-like for the purpose of analysis if the ratio $q_6q_6(i,j)/\sqrt{q_6q_6(i,i)q_6q_6(j,j)}$ was greater than $0.7$.\\

Hexagonal/icosahedral order was also measured via the third-order parameter  $w_6$:

\begin{eqnarray}
w_l(i) &=& \sum_{\substack{{m,n,o}\\{{\rm s.t.}\,m+n+o=0}}} \begin{pmatrix}l&l&l\\m&n&o\end{pmatrix}  q_l^m(i) q_l^n(i) q_l^o(i) 
\end{eqnarray}

where parentheses indicate a Wigner 3-j coefficient determined for the $m,n,o$.  $w4,q4$ and $w6$  bond order parameters were found using the \verb|bop| utility \cite{Wang2005}. In all plots, the $q6$ value is calculated over the 12 nearest neighbours $j$ for each particle, in order to have the most sensitive probe of hexagonal order.\\

The progress coordinate of the flatPERM-FFS calculation was chosen as the number of solid-like bonds (over a threshold of $0.75$) in the largest cluster of particles having 10 or more such bonds.  The initial interface $A$ was defined as $20$ bonds, with further interfaces placed not less than $20$ bonds apart.  The interface placement algorithm of Kratzer et al.~\cite{Kratzer2013} was used.  Only one trajectory per FFS calculation was analysed, in order to avoid complicating the calculation of errorbars by the use of partly correlated trajectories. The University of Luxembourg HPC service was used \cite{Varrette2014}. 
\\

In order to analyse the local environment of particles from an additional perspective, the parameter-free algorithm of van Meel et al.~\cite{vanMeel2012} was used to estimate the number of actual neighbours per particle, with a value of 12 strongly indicating either icosahedral or close-packed order.  This algorithm operates by associating a solid angle to each additional neighbour moving outwards from $i$ until the solid angle subtended is $4\pi$.
\\

\begin{figure}[!ht]
\begin{center}
\includegraphics[width=\columnwidth]{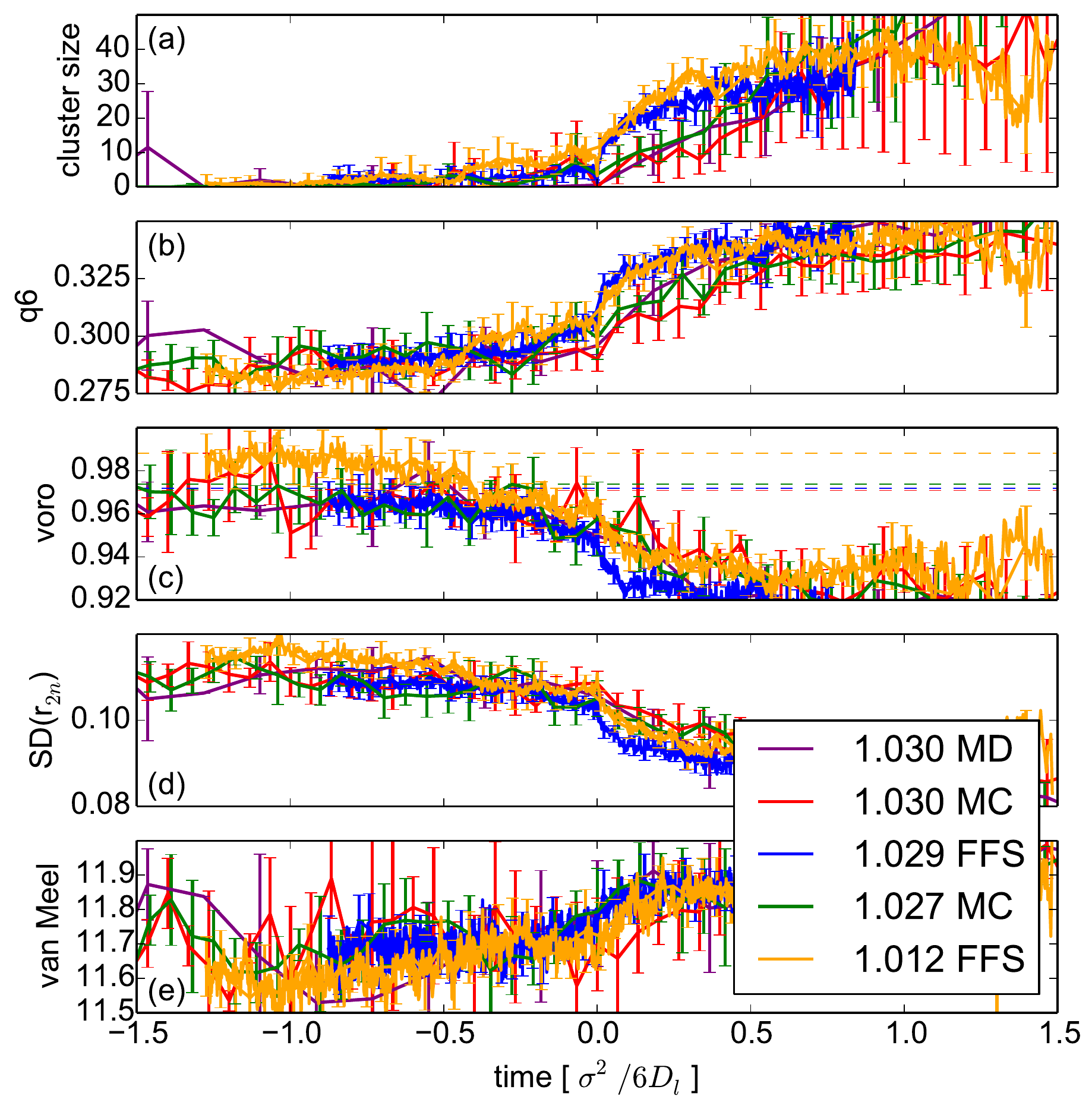}
\end{center}
\caption{\label{fig:statsInBox} ({\bf a}) Number of particles in the largest crystalline cluster present, ({\bf b}) $q6$ hexagonal order statistics ({\bf c}) Voronoi volume, ({\bf d}) per-particle translational order given as the standard deviation of neighbour distances within the first and second shells,  and ({\bf e}) the number of neighbours found using the van Meel algorithm.  Averages are over a sphere of radius 2$\sigma$ (see text). Dashed lines show the whole-system average Voronoi volume. Errorbars are twice the estimated standard error.}
\end{figure}

In fig.~\ref{fig:statsInBox}(a) we show the average size of the largest crystalline cluster. The definition of solid particles for the purpose of plotting this figure was chosen to be the same as in ref.~\cite{Schilling2010}, such that particle $i$ was defined as solid if 11 or more neighbours $j$ gave a $q6q6$ product $>0.7$. This differs slightly from the thresholds chosen to define cluster membership in order to construct a reaction coordinate for the FFS runs, of 10 and $0.75$\label{fn:solid}. Insensitivity of this statistic to threshold choice is shown in the supplementary data (fig.~S1 \cite{suppIbid}) by presenting the same statistic for a minimum cluster size of 7, as in \cite{Russo2012}).  Because nucleation events are randomly distributed in time, it was necessary to define a different time-zero for each trajectory such that the time series collected over different nucleation events could be averaged together.  A space-time zero marking the centre and start of each nucleation event was defined.  The position of the nucleation event was defined as the centre of mass of the largest cluster in the system at the final timepoint when this cluster was of size 30 particles (i.e.~at the time after which it grew irreversibly). The time-zero for the nucleation event was then defined by following the 30 particles backwards in time until all of them were liquid-like. The time zero is thus defined as the {\it first final solidification} of any particle in this cluster. Averaged quantities were then taken over particles within a sphere of radius $2\sigma$ around the nucleation centre.
\\

Across the traces of fig.~\ref{fig:statsInBox}(a-d) we observe a gradual, tandem appearance of densification and ordering.  The same pattern is evident in the additional new datasets as in the reanalysis of the  old data.  We attempt to make a local measurement of translational order distinct to orientational order by showing the standard deviation of the distance to nearest neighbours $SD(r_{2n})$ over the first two shells for each particle \ref{fig:statsInBox}(d). We make this direct-space measurement of translational order because the usual definition of translational order as relating to the presence of higher order Bragg peaks is inherently non-local and therefore difficult to apply to a small region of emerging solid phase, as well as being unnecessary in the case that direct-space information is available.  Using this information we can see that translational and rotational ordering are synchronised, thus eliminating a candidate two-step nucleation mechanism.  By measuring the number of particles in the first neighbour shell via the algorithm of van Meel et al.~\cite{vanMeel2012} we show that the increase of density and order arises by an increase towards 12 in the average number of immediate neighbours \ref{fig:statsInBox}(e), rather than by isotropic compaction or by expulsion of a thirteenth neighbour. 
\\

All observables apart from the size of the largest discrete solid-like cluster start to smoothly deviate from their liquid values at $t\approx-\sigma^2/6D_l$, where $D_l$ is the long time self diffusion constant. The cluster size lifts off more sharply and a little later, because it is defined via a discrete thresholding of the $q6$ values. In refs.~\cite{Schilling2010, Auer2001} this threshold was set close to values expected for the bulk crystal. We suggest that the emphasis on densification as the leading process in ref.~\cite{Schilling2010} arose mainly as result of this decision to treat crystalline order as a binary quantity, while it is now clear that local ordering varies continuously on the pathway between the fluid and crystalline phases.\\

\begin{figure}[!ht]
\begin{center}
\includegraphics[width=\columnwidth]{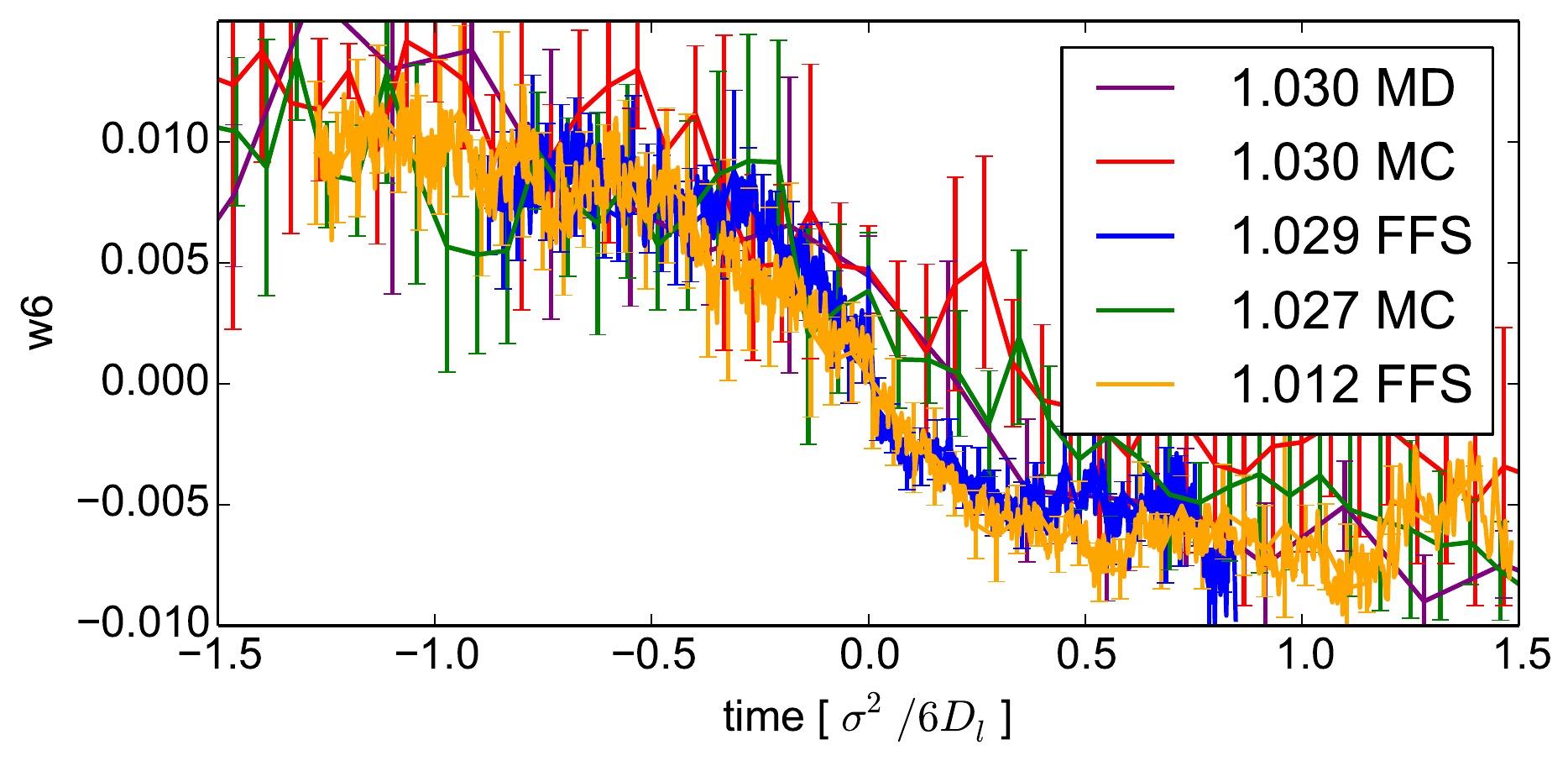}
\end{center}
\caption{\label{fig:w6_trace} The $w6$ spherical harmonic order parameter, used to identify icosahedral structure, does not show sufficient change (i.e.~$w6 < -0.023$) to indicate that icosahedral order plays a role in nucleation at number densities $\phi=[1.012..1.03]$.}
\end{figure}

A neighbourhood of twelve particles is compatible with the (non-spacefilling) icosahedral symmetry as well as with fcc and hcp.  There has been considerable discussion, especially in relation to glass formation, of the idea that this can lead to  a multistep or arrested nucleation process, for a review see \cite{Tarjus2005}.  A small decrease in $w6$ (together with a large increase in $q6$) is associated with hexagonal ordering, however a larger decrease in $w6$ (together with a small decrease in $q6$) signifies icosahedra.  An example threshold from the literature for icosahedral order is $w6 < -0.023$ \cite{Leocmach2012}.  By computing the $w6$ parameter as well as the $q6$ measurements we show that there is no signal of icosahedral order as an aspect of the nucleation process at the non-glassy overcompressions studied here (fig.~\ref{fig:w6_trace}).  
\\

\begin{figure}[!ht]
\begin{center}
\includegraphics[width=\columnwidth]{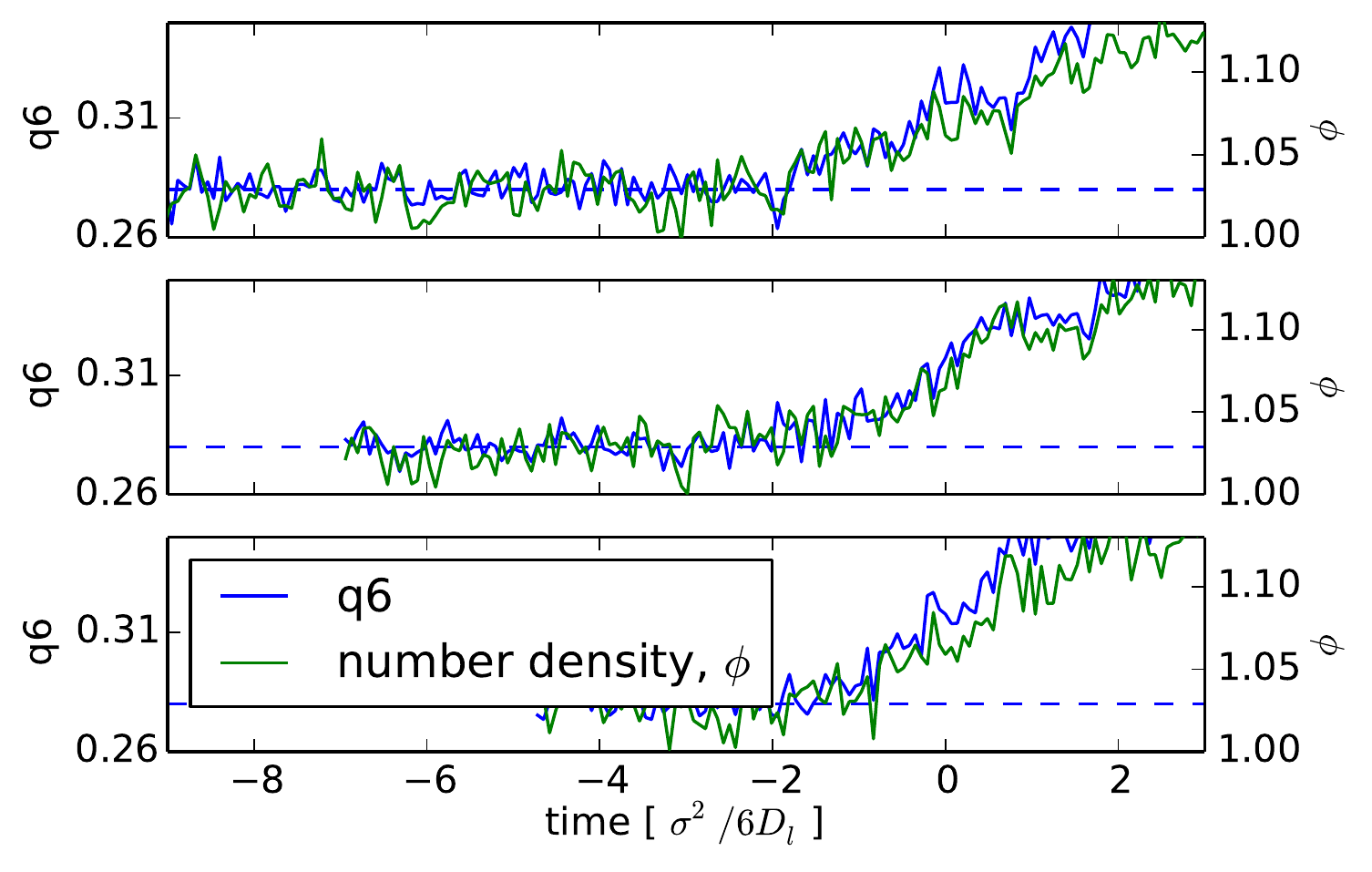}
\end{center}
\caption{\label{fig:statsInBoxTraces} Time series of average $q_6q_6(i,i)$ and number density (inverse Voronoi volume, $\phi$) for three MC runs at $\phi\sigma^3=1.027$, with averages collected in a sphere of radius $2\sigma$ centred on the nucleation event at $t=0$. Order and density increase simultaneously in each case, and fluctuations in the order and density traces are correlated with no time lag, see in particlar the upward-trending parts of each trace. The dashed line shows $q6=0.28$.}
\end{figure}

 To further examine the time-correlation of the density and order changes, we plot single traces (the first three MC runs made at $\phi\sigma^3=1.027$, fig.~\ref{fig:statsInBoxTraces}).  In the same way as for the average pathway, for individual runs the density and order increase simultaneously. Fluctuations evident in the individual traces, whether up or down, are correlated between the two observables with no discernible time lag. From this we confirm the suggestion of fig.~\ref{fig:statsInBox} that the increases of density and order are consequences of the exact same particle displacements, in contrast to the constant-density ordering which has been hypothesized \cite{Russo2012}.\\  

\begin{figure}[!ht]
\begin{center}
\includegraphics[width=\columnwidth]{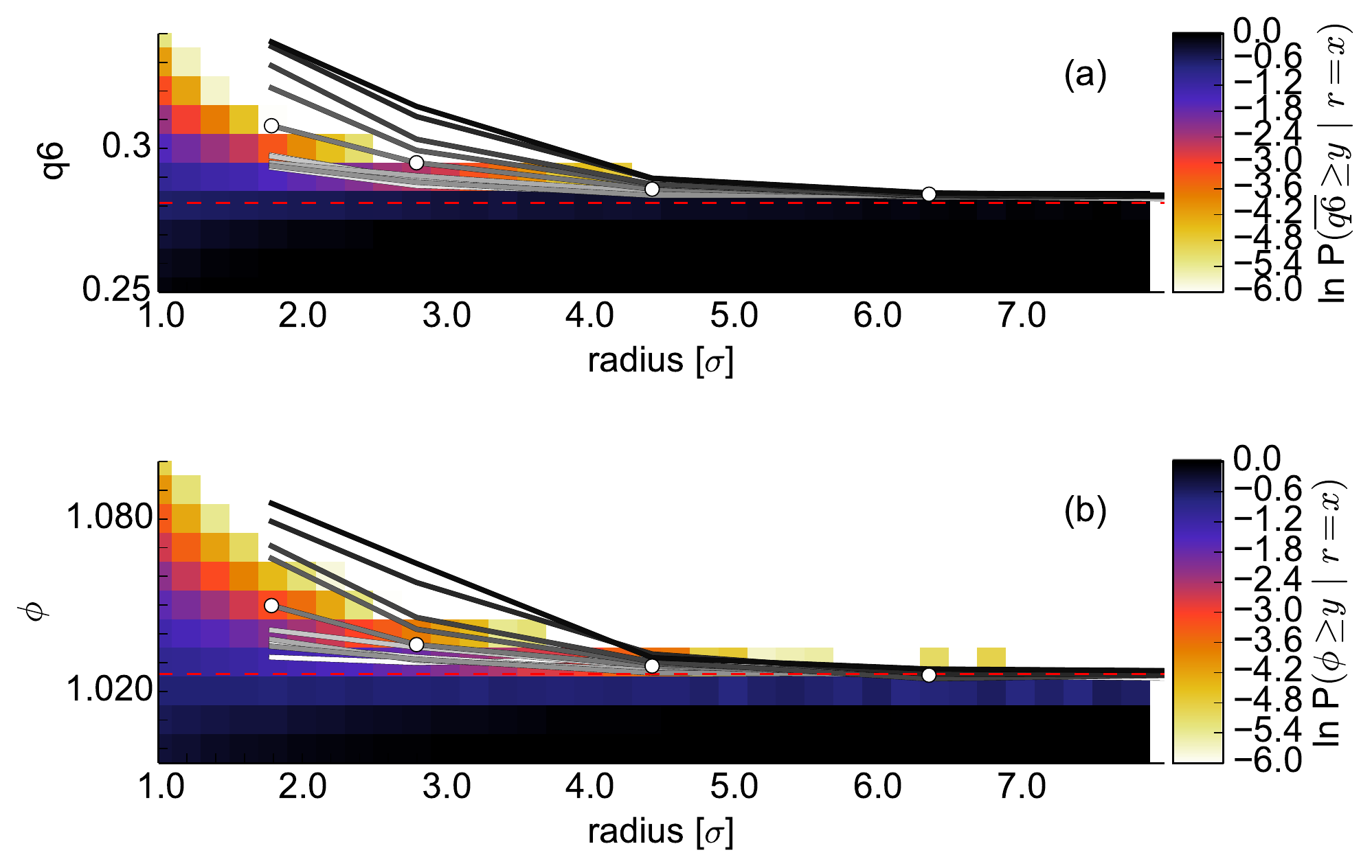}
\end{center}
\caption{\label{fig:vvol_q6_boxL} {\bf(a)} $q6$ and {\bf(b)} inverse Voronoi volume, $\phi$, averaged over spherical shells of increasing radii. The colour maps show the log probability to find a given order parameter value (or greater) when averaging over random shells of the same radius in the bulk fluid. Traces indicate orientational averages over shells centred on the time-zero of the initiation of the nucleation event ({\it first final solidification} of a particle). The trace shading from light to dark indicates a time range of $\sigma^2t/6D_l=-1..1$, with the $t=0$ trace highlighted using circles. The $\phi\sigma^3=1.027$ MC dataset is shown here. Dashed lines indicate quiescent averages.}
\end{figure}

In order to estimate the length and entropy scales associated with the initiating fluctuation, we plot the $q6$ and number density averages leading up to nucleation for the set of MC simulations at $\phi\sigma^3=1.027$, with averages collected over spherical shells of increasing size around the point of initiation of the nucleus, defined as in fig.~\ref{fig:statsInBox} (fig.~\ref{fig:vvol_q6_boxL}, lines). For comparison we also map the log probability of observing a value higher than a given $q6$ or $\phi$ in the bulk liquid, i.e.~the cost in free energy associated with creating an order parameter fluctuation of this magnitude and radius. The trace at first final appearance of the solid-like nucleus (fig.~\ref{fig:vvol_q6_boxL}, line with open circles) is associated with a free energy cost of approximately $3$ to $5\,k_BT$. Gradual and simultaneous densification and ordering is indicated, with ordering initiated non-locally in a (potentially irregular) region of approximately $3-5\sigma$ radius. That the initiating density fluctuation does not become negative over the range plotted indicates that the inward mass flux is compensated at fairly long range. A version of fig.~\ref{fig:vvol_q6_boxL} confirming similar behaviour at $\phi\sigma^3=1.012$ is appended as supplementary data (fig.~S2 \cite{suppIbid}).\\

\section{Conclusion}

We have analysed the early stages of crystal nucleation in hard spheres. We observe that densification does not occur prior to bond orientational ordering, and also that bond orientational ordering does not occur prior to densification. Hard sphere nucleation starts with a process that includes densification and ordering. As far as we are able to measure orientational and translational ordering as distinct phenomena, we find that they also occur simultaneously.\\

 The early-stage hexagonally ordered fluctuations which we discuss here are identifiable with the `dense amorphous regions' or `low symmetry clusters' used by Schilling et al.~\cite{Schilling2010} to argue the density-first case in 2010, and also appear at least similar to the `precursor structures' suggested to form without density enhancements by the Tanaka group \cite{Kawasaki2010a,Kawasaki2010b,Russo2012} and used to argue an order-first position. The explanation which we can offer for the historical divergence between density-first and order-first opinions is that probes of differing sensitivity to the respective phenomena have previously been employed. By showing that both variables lift from their quiescent distributions not only together on average, but with instantaneously correlated fluctuations in individual traces, we hope that we have settled this dispute.\\

We assert that nucleation of hard sphere crystals at low to medium overcompression begins with a collective fluctuation of radius approximately $4\sigma$, simultaneously manifested in the density and in the positional order. The radius of $4\sigma$ corrresponds to order 100 particles (depending on the shape of the nascent dense ordered region), indicating fluctuations of a highly collective nature. The increase of density at the nucleation site is supplied by long-range collective mass transport, with the amplitude of both density and order falling off gradually over the characteristic distance.\\

We accept the current speculation that it is possible to advance a description of the hard sphere nucleation process as two-step, but only in the weak sense that it begins with an order/density fluctuation which initially strengthens more than it grows (before starting to grow more than it strengthens).  At least when away from the glassy regime, the changeover between these two `steps' is smooth and there is no sign of any qualitative changes in the type of order manifested.\\

The most agreement that we can find with and across the recent literature is that nucleation is initiated with a diffuse entity of finite spatial extent: this model contrasts with the na\"ive image of a sharply defined region of daughter phase expanding outwards from a point, and is a good candidate to eventually supply control and understanding of nucleation phenomena by matching the lengthscale and structure of experimental probes with those of the initial fluctuation.\\

\begin{acknowledgments}
This paper is dedicated to the memory of Ian Snook.  We thank H.~Sch\"ope, M.~Oettel, J.\ Lutsko and H.~Tanaka for fruitful discussions.  Authors SD and MA were supported by grant FRPTECD of the Fonds National de la Recherche Luxembourg.
\end{acknowledgments}


%

\end{document}